# Ubuntu One Investigation: Detecting Evidences on Client Machines

Mohammad Shariati[1]; Ali Dehghantanha[1]; Ben Martini[2]; Kim-Kwang Raymond Choo[2]

[1] School of Computing, Science and Engineering, University of Salford, Greater Manchester, United Kingdom

[2] University of South Australia, GPO Box 2471, Adelaide, SA 5001, Australia

m_shariati83@yahoo.com; A.Dehghantanha@salford.ac.uk; Ben.Martini@unisa.edu.au; Raymond.Choo@unisa.edu.au

## Abstract

STorage as a Service (STaaS) cloud services have been adopted by both individuals and businesses as a dominant technology worldwide. Similar to other technologies, this widely accepted service can be misused by criminals. Investigating cloud platforms is becoming a standard component of contemporary digital investigation cases. Hence, digital forensic investigators need to have a working knowledge of the potential evidence that might be stored on cloud services. In this chapter, we conducted a number of experiments to locate data remnants of users' activities when utilising the Ubuntu One cloud service. We undertook experiments based on common activities performed by users on cloud platforms including downloading, uploading, viewing and deleting files. We then examined the resulting digital artefacts on a range of client devices, namely Windows 8.1, Apple Mac OS X and Apple iOS. Our examination extracted a variety of potentially evidential items ranging from Ubuntu One databases and log files on persistent storage to remnants of user activities in device memory and network traffic.

## 1. Introduction

The term cloud computing refers to a model whereby a user can access computing resources via a network on an on-demand basis (Mell and Grance, 2011). Various types of resources can be shared between users and in a way that remote clients can utilise them e.g. processing, volatile and persistent storage and so on. This pool of resources is commonly available as a service via an internal network (private cloud) or publically via the Internet (public cloud). In addition to providing the de facto definition of cloud computing the National Institute of Standards and Technology (NIST) also defined a number of service models including:

Software as a Service (SaaS), Platform as a Service (PaaS) and Infrastructure as a Service (IaaS) (Mell and Grance, 2011). Storage as a Service (STaaS) is an addition to these traditional service models. STaaS technologies enable users to store, download and share their data in a very accessible manner. There are a number of STaaS service providers including Dropbox, Microsoft OneDrive, Google Drive and Ubuntu One. These service providers commonly provide personal accounts for minimal or no cost. Cloud Service Providers (CSPs) have made significant efforts to attract customers by supporting various types of devices ranging from traditional PC platforms such as Windows, Mac OS X and Linux to more recent smart phone operating systems such as iOS and Android. Also, CSPs generally offer access to their services via standards compliant web browsers including Internet Explorer, Google Chrome, Mozilla Firefox and Apple Safari. These features allow users to access their data via the majority of internet connected devices.

However, while STaaS services provide legitimate users with significant utility and convenience, they are equally useful to criminals who utilize them for storing and sharing illicit materials. The global nature of cloud computing infrastructure contributes to the numerous technical and jurisdictional challenges in the identification and acquisition of evidential data by law enforcement and national security agencies.

Digital forensics is the process of identifying, preserving, analysing and presenting evidence for use in legal proceedings (McKemmish, 1999). The process of traditional forensic investigation is often impeded by some of the key characteristics of the cloud environment such as multitenancy and global data distribution. Taylor et al. (2011) highlighted that with the advent of cloud computing acquiring and analysing digital evidence from cloud services using traditional processes is generally infeasible. One key area of difficulty is in identifying the particular service utilized by suspects and then extracting potential remnants of user activities involving that service.

In this chapter we seek to assist forensic investigators and practitioners to detect possible evidential remnants derived from the Ubuntu One cloud storage service. The artefacts discussed in this chapter should assist in detecting the use of Ubuntu One and the associated evidential remnants stored on client devices. The focus of this study is to detect file system, RAM, and network artefacts present after utilizing Ubuntu One on the Windows 8.1, Mac OS X 10.9 and iOS 7.0.4 platforms.

In this chapter, we intend to address the following research questions:

1. What data can be found on a device's persistent storage after using the Ubuntu One client software and the location of data remnants within Windows, Mac OS X and iOS devices?
2. What data can be found in a device's persistent storage after using Ubuntu One via a web browser?
3. What data can be extracted from volatile memory on Windows and Mac OS X devices when utilising Ubuntu One?
4. What data can be extracted from collected network traffic after Ubuntu One has been accessed on Windows, Mac OS X and iOS devices?

The remainder of this chapter is organized as follows; in the next section we provide a brief review of related work in the field of cloud forensics. In section 3 and 4 we outline both the methodology and experiment setup utilised in our experiments, respectively. In section 5 we present our research findings and finally in section 6 we conclude the chapter.

2. **Related Work**

Grispos, et al. (2012) discussed a number of challenges for forensic investigations in the cloud, namely creating valid forensic images, recovery of distributed evidence and management of large data sources. There are a number of other research studies that highlight a number of the major issues of cloud forensics (Biggs and Vidalis, 2009; Birk and Wegener, 2011; Martini and Choo, 2012, M. Damshenas et al., 2012, F.Daryabar et al., 2013, A. Aminnezhad et al., 2013).

In the case of STaaS forensic research, the majority of existing research has been conducted on STaaS clients, with a smaller subset of the published materials focusing on server side STaaS investigation. Quick and Choo (2013a, 2013b, 2013c, 2014) have developed a forensic framework to identify, acquire and present evidential data remnants of Dropbox, Google Drive and Microsoft SkyDrive on the Windows 7 and iPhone platforms. Hale (2013) published a similar investigation on the Amazon Cloud Drive client on Windows XP and 7. In addition, Chung et al. (2012) analyzed Amazon S3, Google Docs and Evernote and outlined a technique to collect data from personal computers and mobile devices. Federici (2014) described the concepts and internals of the Cloud Data Imager tool which he developed to provide read only access to files and metadata of selected remote folders on

STaaS services and currently provides access to the Dropbox, Google Drive and Microsoft SkyDrive services. In terms of server STaaS analysis, Martini and Choo (2013) focused upon the client and server artefacts created with use of ownCloud. The analysis of the ownCloud server component, after analysis of the client component, allows the practitioner to obtain a wider range of evidential data (e.g. previous versions of files).

The numerous publications that investigate STaaS products demonstrate the need for researchers to undertake detailed analysis to guide practitioners in collecting all available evidence from cloud storage products.

## 3. Methodology

Using Ubuntu One as a case study, artefacts were identified that are likely to remain after the use of cloud storage, in the context of several experiments conducted on Windows, Mac OS X and Apple iPhone 3G clients. As Ubuntu One supports accessing, uploading and sharing data using both client software and a browser, we have undertaken experiments across multiple platforms to locate evidential data sources on different client devices.

In each experiment, the investigator first determines whether it is possible to collect volatile data on the platform being investigated. If so, the investigator acquires the contents of physical memory and captures the network traffic. Next, if non-volatile data can be obtained, the investigator gathers data from the file system such as log files, configuration files, internet history data, databases and directories. For the Windows and Mac operating systems, we were able to collect volatile and non-volatile data, but in the case of iOS only network traffic was collected. This was due to the lack of opportunities for forensically sound physical memory acquisition on iOS devices. After collection the investigator, searches for traces of the Ubuntu One cloud storage service in the collected images.

## 4. Experiment setup

The research experiment was broken into six stages namely: 1) preparing the Virtual Machines and iPhone, including installing the cloud applications; 2) uploading a data set to the cloud storage provider; 3) accessing the data through the client application/web browser on the VMs and iPhone; 4) perform various file manipulations to the data set on both the VMs and iPhone; 5) process the VMs and iPhone to extract volatile and non-volatile data and

6) use numerous forensic tools to analyse the collected forensic images and present the final result.

We undertook experiments within the following four usage environments:

 1) Windows browser-based (see Section 5.1);

 2) Windows app-based (see Section 5.2);

 3) Mac OS X app-based (see Section 5.3); and

 4) iOS app-based (see Section 5.4).

We have used Ubuntu One 4.0.2 which provides users with 5GB of free space and utilized the following three files from the Enron email dataset, downloaded from the project website (http://bailando.sims.berkeley.edu/enron_email.html) on 15$^{th}$ of April 2014, to conduct our experiments:

1. AQUA-OS2.BMP (151KB),

2. HANGING.DOC (22KB),

3. HANGING.txt (2KB)

Different file types were utilised in the experiments to determine whether any discrepancies in forensic collection were observable for the different file types.

Windows 8.1 and Mac OS X 10.9 experiments were conducted on virtualized environments utilizing VMware Player 6.0.2. An iPhone 4S with iOS 7.0.6 was used to undertake iOS investigation experiments. Each VM was configured with one CPU, 2GB of RAM and 20GB of hard disk space.

Our experiments were designed to simulate common user activities on cloud platforms namely uploading, downloading, opening and deleting files. For the purposes of this research, one set of credentials were used in all experiments to simplify the location of the credentials as part of forensic image examination.

**Web Browser Investigation Experiments Setup**: To commence our environment setup for web browser analysis we installed the four most popular browsers at the time of research, namely Microsoft Internet Explorer (version 10.0.9200.16384), Mozilla Firefox (version 25.0.1), Google Chrome (version 31.0.1650.63), and Apple Safari (version 5.1.7), on four VMs. We then performed a series of upload, open, download and delete operations with one

VM for each type of operation. Table 1 outlines the list of tools that were copied to each VM for monitoring changes and detecting possible evidential data.

Table1. Software used on VMs for analysis

| Software | Version | Purpose |
|---|---|---|
| Regshot | 1.9.0 | Registry Monitor |
| Process Monitor | 3.05 | Process, Registry and File Activity Monitor |
| Nirsoft web browser passview | 1.43 | Saved Password Retrieval |
| Digital Detective Net Analysis | 1.5 | Browser Cache Retrieval |

Figure 1 shows the VM hierarchy for our browser based experiments.

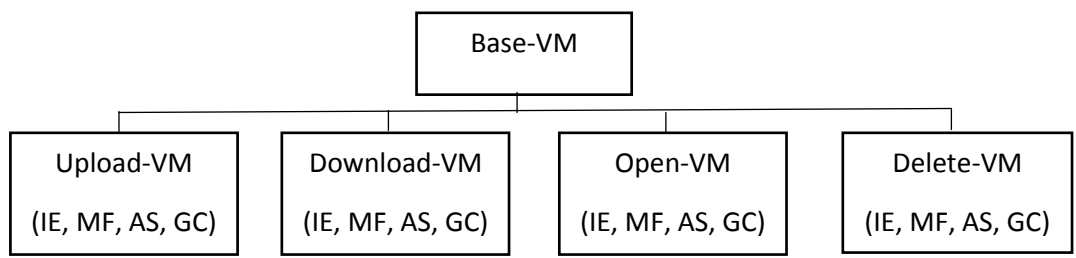

Figure 1. VMs created for web browser usage investigation

**Windows App-Based Investigation Experiment Setup**: In addition to the series of upload, download, open and delete operations, we also experimented installing and uninstalling the Ubuntu One app to determine the artefacts that could be detected after such activities on client devices, as outlined in Figure 2.

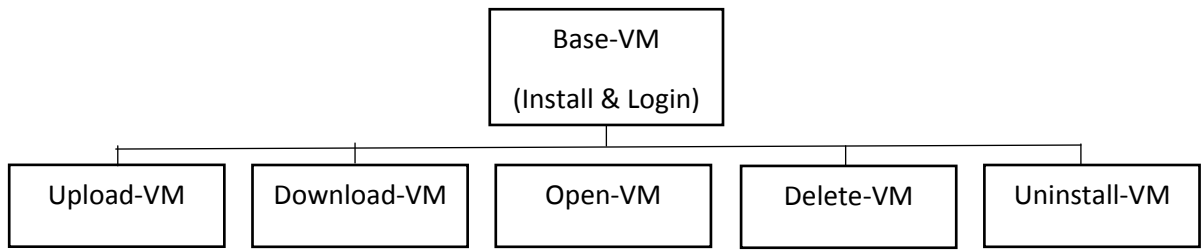

Figure 2. VMs created for Windows and Mac OS X platforms investigation

**Mac OS X App-Based Investigation Experiment Setup:** Similar experiments to Windows platform were conducted on Mac OS X 10.9 Mavericks. However, the process for uninstalling applications on Mac OS X is different in comparison to uninstalling an app in the

Windows environment. While most Windows programs include an uninstaller that can be run using the Add/Remove Programs tool available in Control Panel, no such feature exists in Mac OS X and therefore, most users simply move application bundles to the Trash. Assuming that the trash is not emptied, we should be able to locate significant application artefacts after an uninstallation of the Ubuntu One client on OS X client machines.

**iOS App-Based Investigation Experiment Setup:** For our iOS experiment we used a jailbroken iPhone 4S running iOS 7.0.6 to conduct our experiments. iFile 2.0.1-1 was installed from Cydia to browse iOS storage directories. The directory that holds the associated files and folders of apps from the Apple Store is /private/var/mobile/Applications, with each app being assigned a subdirectory name by universally unique identifier (UUID). Activating the 'Applications Names' option under iFile's Preferences (File Manager section) translates the UUIDs to the human readable names of all the subdirectories. Mobile Terminal was another application installed from Cydia enabling execution of UNIX commands in the iOS environment.

## 5. Discussion and Analysis

In this section, we explore residual artefacts generated by Ubuntu One when cross-platform methods are used to manipulate data hosted on the cloud. Generally, collection of data remnants was conducted in two stages. The first stage is the acquisition of live data. The analysis of this live (volatile) data is regarded as a significant in recovering sensitive information that is available while Ubuntu One is being accessed. The second stage is the analysis of non-volatile data remnants that can be located on the local system. These two methods complement each other to maximise the amount of available evidential data during an investigation. In our research, forensic analysis of live data encompasses analysis of live memory and network traffic, while data remnants analysis involves the persistent files such as log files, databases and the registry (for Windows platforms).

In all versions of the client applications, after lunching Ubuntu One, the user must enter a device name and their authentication credentials. At next launch, Ubuntu One logs-in automatically unless the user unlinks the device.

For this research, virtual hard disks, virtual memory, and forensic images of real memory and network traffic were examined using multiple forensic tools. We analysed the VM's VMEM file as a memory dump file and the VMDK file as an image of the hard disk using AccessData FTK (version 1.86.1). We also used Hex Workshop (version 6.7) for analysis of memory and hard disk images, which enabled searching for keywords such as the Ubuntu One credentials, files being accessed and words such as "Ubuntu One", "Ubuntu" and "UbuntuOne". The network traffic was captured and analysed using Wireshark (version 1.10.2) and further analysis was conducted using NetworkMiner (version 1.5). SQLite DB Browser under OSForensics (version 2), PList Explorer (version 1.0) and Notepad++ (version 6.4) were employed to access and retrieve evidential data from the Ubuntu One databases and log files. AccessData Registry Viewer (version 1.7.4.2) and Regripper (version 2.8) were utilised to analyze Windows registry and NTUSER.dat files.

## 5.1 Windows browser-based

Ubuntu One allows users to access and manipulate their data on the cloud without installing the client application, via a web-browser. From an evidence collection perspective, it can be presumed that the browser-based application leaves fewer remnants on local computer compared to the full client application. The following stages outline the results of our analysis, in a step-by-step manner, for each of the evidence sources identified such as live memory and browser cache.

### 5.1.1 Memory

We found live memory forensic analysis very useful for extracting important digital artefacts when Ubuntu One was being accessed via the web interface. We utilised two methods to detect Ubuntu One user identity information in the live memory:

Method 1- Searching for the string "login&password=" to retrieve the user's credentials in plain text (see Figure 3):

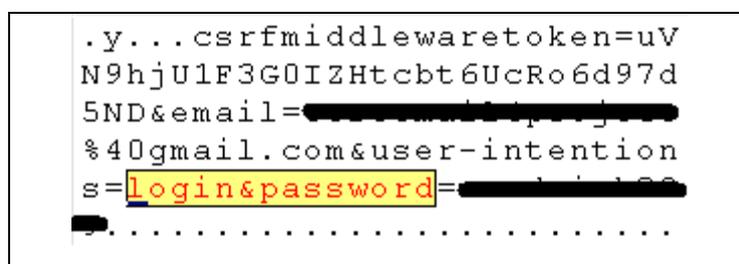

**Figure 3. User credentials located in live memory**

Method 2- Searching for the string "openid.ax.value.email" to retrieve the users email address (see Figure 5):

```
name="openid.ax.value.email
.1" value="                
t@gmail.com"/><input type="
```

Figure 5.Username located in live memory

We also noted that it was possible to extract the names of files that had been accessed or manipulated depending on the specific operation used. A selection of operations and the associated artefacts are outlined in Figures 6 to 8.

```
.    .    <li>.        <a ti
tle="Delete File" alt="Dele
te File".        id="fus
uENo_p5wTZCKgkT9NNi2YQ-dele
te".            class="ul-fi
les-delete-link delete-butt
on">Delete file</a>.    </l
i>.     </ul>.    </td>.</tr
>...    .            .<
tr id="fusWbdqbNWeQnu8YNOUG
bheow".    class="file".
 title="File">.    <td clas
s="files-td-name" id="HANGI
NG.txt">.        .
    <a title="HANGING.txt"hre
f="https://files.one.ubuntu
.com/WbdqbNWeQnu8YNOUGbheow
" target="_blank">HANGING.t
```

Figure 6. Filename located after delete operation

```
......?.....Content-Length:
 0..Content-Disposition: at
tachment; filename=HANGING.
txt..Vary: Accept-Encoding,
```

Figure 7. Filename located after open/download operation

```
ata; name="file"; filename=
"HANGING.txt"..Content-Type
```

Figure 8. Filename in upload operation

**5.1.2 Browser cache and history**

When a user has accessed Ubuntu One via the online interface the web browsers cache and history may contain evidential data and should be extracted. Although it will not generally be possible to extract the Ubuntu One user credentials from the browser cache and history, numerous other evidential artefacts can often be retrieved. While we were not able to extract credentials from the cache directly, in the case of the Chrome and Internet Explorer browsers, the Nirsoft Web Browser Pass View was able to extract the stored password which we saved using the browser.

For an investigator, the first step in a cloud investigation is often to determine which cloud storage services have been used by the suspect, and URL addresses are one source of evidential data useful in determining this. In our experiments, we noted a number of web addresses in the cache and history that relate to Ubuntu One. These addresses are listed below:

https://media.one.ubuntu.com

https://one.ubuntu.com

https://login.ubuntu.com/

https://files.one.ubuntu.com/

From our analysis of the cache data we determined that there were three web pages that we considered to be of particular importance, namely "dashboard.htm", "files.htm" and "+opened.htm".

The dashboard.htm file contains the first name and last name that has been entered by the Ubuntu One user. In our case, "Test" and "Project" were the first name and last name respectively (see Figure 9).

Figure 9. The users full name in dashboard.htm

The opened.htm file stores the username as well as the full name of the Ubuntu One user (see Figure 10).

Figure 10. The users username and full name in opened.htm

Finally, files.htm cache files often contain filenames, file size and the date and time that an operation on the file was carried out (see Figure 11).

```
<td class="files-td-name" id="AQUA-OS2.BMP">

    <a title="AQUA-OS2.BMP" href="https://files.one.ubuntu.com/YUS8opQZQSOCavYpI-3sDQ" target="_blank">AQUA-OS2.BMP</a>

</td>
<td class="files-td-size">

        150.1 KB

</td>
<td class="files-td-date">

            <span title="2014-04-22 03:15:30">2014-04-22</span>

    </td>
```

Figure 11. Filename and associated timestamp in files.htm

### 5.1.3 Registry

There was no information regarding Ubuntu One credentials and usage located in the registry except for a "TypedURL" entry from Internet Explorer (see Figure 12). As the name suggests, these registry entries are stored when a user types a URL in Internet Explorer (Mee, Tryfonas & Sutherland 2006).

```
TypedURLsTime
Software\Microsoft\Internet Explorer\TypedURLsTime
LastWrite Time Tue Apr 22 03:48:04 2014 (UTC)
    url1 -> Tue Apr 22 03:48:04 2014 Z (http://one.ubuntu.com/)
```

Figure12. "TypedURL" in the Windows registry

### 5.1.4 Network traffic

Wireshark was used for collecting the network traffic from Ubuntu One usage, which was then analysed using NetworkMiner. We found that all of the collected traffic was encrypted due to the use of SSL/TLS when communicating with the Ubuntu One servers. For this reason we were unable to extract any data of significant evidential value. However a number of common IP addresses were used for Ubuntu One communications, which resolve to Ubuntu.com subdomains. Table 2 shows the IP addresses and associated hostnames that were extracted from the network traffic capture file after a login to Ubuntu One and contents of the user's account had been accessed/manipulated.

Table 2. List of IP addresses and hostnames extracted from network traffic.

| IP Address | Hostname |
| --- | --- |

| 91.189.89.77 - 91.189.89.78 | one.ubuntu.com |
|---|---|
| 91.189.89.182 - 91.189.89.183 | media.one.ubuntu.com |
| 91.189.89.206 - 91.189.89.207 | login.one.ubuntu.com |

## 5.2 Windows app-based

Upon installation the Ubuntu One client software creates a folder named Ubuntu One stored in the "\Users\<user>\Ubuntu One" directory, by default. The folder only appears to be utilised while the Ubuntu One client is running, during which content can be found in the directory as discussed below. By default, this folder is used by Ubuntu One for automatic synchronisation of files. In addition to this directory, the installation of Ubuntu One creates some folders on the local computer to store persistent data including log files, databases and other related files. Utilizing Process Monitor, we detected the following folders were used by Ubuntu One:

1) C:\ProgramData\Microsoft\Windows\Start Menu\Programs\Ubuntu One

2) C:\Program Files (x86)\ubuntuone\

3) C:\Users\[user]\AppData\Local\ubuntuone

4) C:\Users\[user]\AppData\Local\xdg\cache

5) C:\Users\[user]\AppData\Local\xdg\ubuntuone

6) C:\ProgramData\ubuntuone

7) C:\Users\[user]\Ubuntu One

We also noted three processes related to the use of the Ubuntu One client namely *ubuntu-sso-login.exe*, *ubuntuone-control-panel-qt.exe*, *ubuntuonesyncdaemon.exe*.

## 5.2.1 Memory:

Unlike our browser-based experiments discussed above, analysis of live memory for the Windows client did not result in the location of the user's password in plaintext. However the username was located by searching for the string "https://login.ubuntu.com/+id" (see Figure 14):

Figure 14. Located username in client memory

We were also able to locate filenames for files being accessed from the Ubuntu One default folder however this required the use of double escaped backslashes. For example, when the default path is C:\Users\username\Ubuntu One, the term to search for in the image of live memory is "C:\\\\Users\\\\username\\\\Ubuntu One\\\\" (see Figure 15):

Figure 15. Located filenames in client memory

### 5.2.2 File System

The Ubuntu One folder in Program Files contains many files including .pem and .conf files. The .pem files are certificate files in the PEM encoding format and .conf files contain configuration values for the Ubuntu One client. There is only one log file in the folder, namely install.log, which holds information regarding Ubuntu One's installation process.

The Ubuntu One folder in ProgramData also stores configuration files as well log files in .xdg format. We determined that information of importance can be recovered from the "xdg" log files, including the username used for logging into Ubuntu One and the name and path of the files listed below:

C:\Users\[user]\AppData\Local\xdg\cache\sso\sso-client.log

C:\Users\[user]\AppData\Local\xdg\cache\sso\sso-client-gui.log

Surprisingly, no database files were located in Ubuntu One's folders.

### 5.2.3 Event logs

Windows event logs store useful and valuable information about a system and its users (Do et. al 2014). Depending on the enabled logging level and the installed version of Windows, event logs may provide investigators with valuable data about application operations, login timestamps for users and other system events of interest.

In our research, we located several logs associated with Ubuntu One within the Application and Services log. Searching for the keyword "Ubuntu" in the Windows event log leads to several hits as shown in Figure 16.

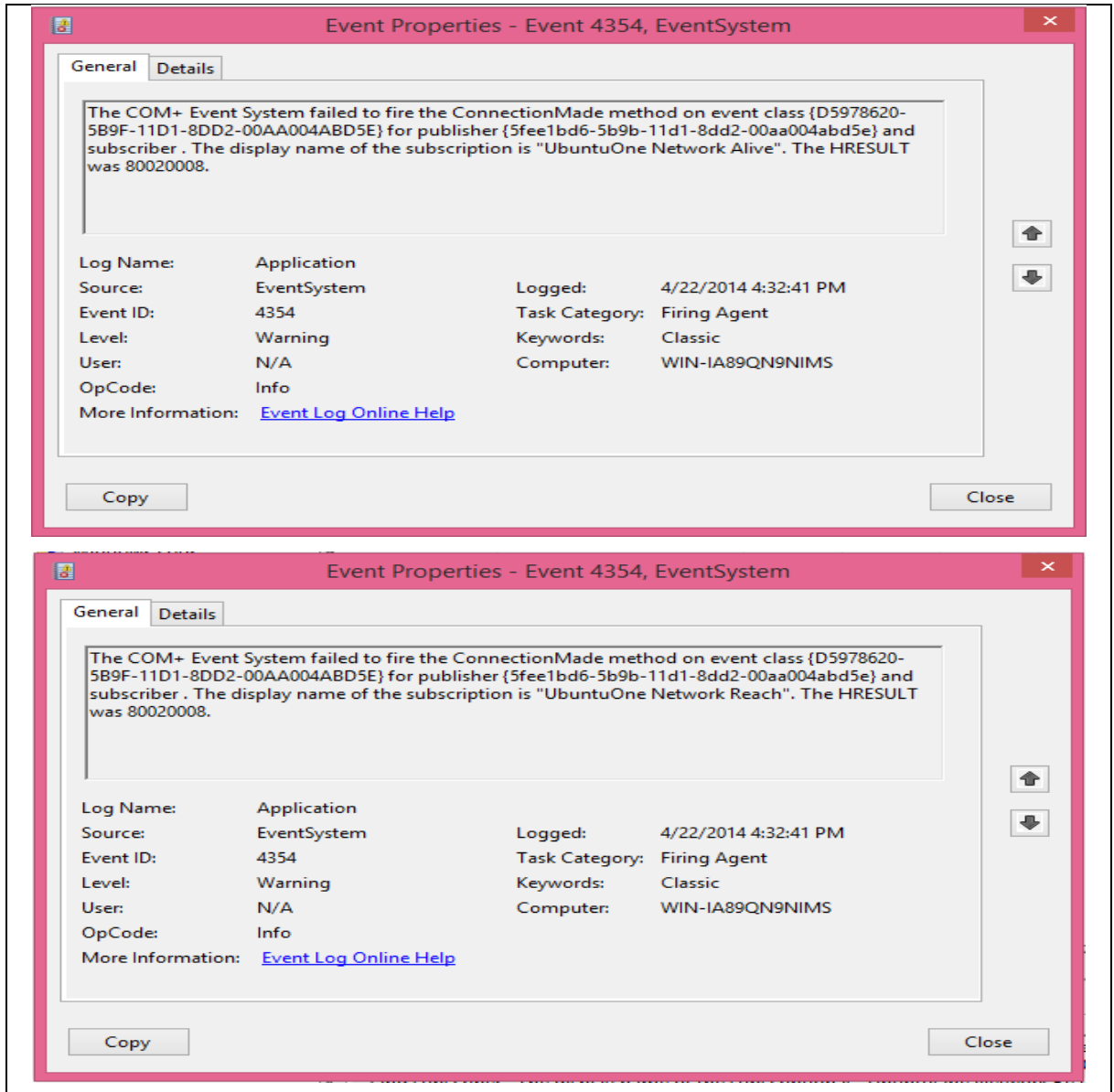

Figure 16. Evidence located in the Windows Event Log

### 5.2.4 Registry

We noted the creation of the following keys in the registry after installation of the Ubuntu One client:

HKEY_LOCAL_MACHINE\SOFTWARE\Wow6432Node\Ubuntu One

HKEY_CURRENT_USER\Software\Ubuntu One

Also, searching for the keyword "Ubuntu One" located the following results in the registry:

HK_Current_User/Software/Microsoft/Windows/Run:

C:\Program Files (x86)\ubuntu one\dist\ubuntuone-syncdaemon.exe

HK_Current_User/Software/Microsoft/Windows/Current version/UFH/SHC:

HKEY_LOCAL_MACHINE\SOFTWARE\Wow6432Node\Microsoft\Windows\Curr
entVersion\Uninstall\Ubuntu One 4.2

Opening files synced via the Ubuntu One client leads to addition of entries in the RecentDocs registry subkey as expected (see Figure 17):

```
recentdocs v.20100405
(NTUSER.DAT) Gets contents of user's RecentDocs key

RecentDocs
**All values printed in MRUList\MRUListEx order.
Software\Microsoft\Windows\CurrentVersion\Explorer\RecentDocs
LastWrite Time Tue Apr 22 11:10:53 2014 (UTC)
   1 = Ubuntu One
   3 = HANGING.txt
   2 = HANGING.DOC
   0 = AQUA-OS2.BMP
```

Figure 17. Recently opening files located in the Windows registry

We did not locate any data regarding the files that had been uploaded or downloaded or Ubuntu One credentials in the registry.

### 5.2.5 Network traffic

Our results for network traffic capture when using the Ubuntu One client was similar to our findings for browser-based access as all network data is encrypted using SSL. As such, plaintext data of value could not be found. However, the captured network traffic shows some differences compared to the network capture acquired using the Ubuntu One online interface. Quick and Choo (2013a, 2013b, 2014) observed Online Certificate Status Protocol traffic relating to the presented SSL certificate in their captured network traffic. We did not note an OSCP query in our network capture. In addition, there were no connections recorded associated with the "media.ubuntu.com" subdomain. The list of IP addresses and associated hostnames extracted from the network traffic collected is listed in Table 3.

Table 3. List of IP addresses and hostnames extracted from the Ubuntu One Windows client network traffic.

| IP Address | Hostname |
|---|---|
| 91.189.89.77 - 91.189.89.78 | one.ubuntu.com |
| 91.189.89.206 - 91.189.89.207 | login.one.ubuntu.com |

**5.2.6 Uninstallation**

The uninstallation process for Ubuntu One removes all of the files located in the Ubuntu One folder in "Program Files" except for the "dist" folder. All other folders associated with Ubuntu One remain on the client machine after uninstallation, including the Ubuntu One default folder and its contents.

Uninstallation also only removes Ubuntu One artefacts from the "HKEY_CURRENT_USER\Software" key, while the remaining registry artefacts are left intact.

**5.3 Mac OS X app-based**

After installation of the Ubuntu One client on Mac OS X 10.9 we located the following directories created by the installation process:

/Applications/Ubuntu One.app

/Users/<user>/Ubuntu One (default Ubuntu One directory)

/Users/<user>/Library/Application Support/Ubuntuone

/Users/<user>/Library/Caches/ubuntuone

/Users/<user>/Library/Caches/sso

Uninstallation of the Ubuntu One client only removes the first directory while the rest remain intact. Some of these directories contained information related to Ubuntu One credentials and sync files such as the following:

~/Library/Caches/sso/sso-client.log

~/Library/Caches/sso/sso-client-gui.log

~/Library/Caches/ubuntuone/syncdaemon.log

The username of the Ubuntu One user can be located within the first two log files as shown in Figure 18:

2014-05-21 20:49:30,353:353.214979172 - ubuntu_sso.current_user_sign_in_page - INFO - CurrentUserSignInPage.login for: ●●●●●●●●●@gmail.com

Figure 18. Username located in OS X Ubuntu One log file

Files which have been synced can be found in the syncdaemon.log log file as can be seen in Figure 19:

2014-05-22 20:24:41,213 - ubuntuone.SyncDaemon.sync - DEBUG - -:-:- - ['-':':'-'] ''/Users/Test/Ubuntu One/AQUA-OS2.BMP'' | EVENT: SV_FILE_NEW:() with ARGS:('',
2014-05-22 20:24:41,213 - ubuntuone.SyncDaemon.sync - INFO - -:-:- - ['-':':'-'] ''/Users/Test/Ubuntu One/AQUA-OS2.BMP'' | Calling new_file (got SV_FILE_NEW:())
2014-05-22 20:24:41,214 - ubuntuone.SyncDaemon.fsm - DEBUG - set_node_id: path='/Users/Test/Ubuntu One/AQUA-OS2.BMP' mdid='37a3fd80-3281-4871-acce-2ba4137ea007'
2014-05-22 20:24:41,214 - ubuntuone.SyncDaemon.fsm - DEBUG - create: path='/Users/Test/Ubuntu One/AQUA-OS2.BMP' mdid='37a3fd80-3281-4871-acce-2ba4137ea007' sha

Figure 19. Synced Filenames located in OS X Ubuntu One log file

### 5.3.1 Memory

We were unable to locate plaintext credentials in our memory capture file (.vmem) of the Ubuntu One client on the Mac OS X platform. However, searching for the Ubuntu One default directory string in memory led to a number of filenames when we conducted upload, download and delete operations (see Figure 20).

...../Users/Test/Ubuntu One
/AQUA-OS2.BMP.............

Figure 20. Filenames of synced files located in Mac OS X Memory

From the other keywords present in the memory capture we were able to determine the operation that was being undertaken on the file. For example, "*ubuntuone.SyncDaemon.EQ - DEBUG - push_event: FS_FILE_CREATE, kwargs: {'path':*" and *"EVENT: FS_FILE_DELETE:{} with ARGS*" represent upload and delete operations respectively.

### 5.3.2 Network traffic

The results of our network traffic capture is similar to that of the Windows client application discussed above, where we did not note any OCSP queries. Table 4 outlines the list of IP addresses (and associated hostnames) extracted from the network traffic capture file.

Table 4. List of IP addresses and hostnames extracted from OS X client network traffic.

| IP Address | Hostname |
|---|---|
| 91.189.89.77 - 91.189.89.78 | one.ubuntu.com |
| 91.189.89.206 - 91.189.89.207 | login.one.ubuntu.com |

## 5.4 iOS app-based

There does not appear to be an official Ubuntu One app available for iOS in the Apple App Store. We selected the "U1Files" (version 0.5) unofficial Ubuntu One client app for analysis on iOS. We installed the app on a jailbroken device running iOS 7.0.4. Application data for apps installed from the App Store is stored in /var/mobile/Applications on iOS. The iFile app (version 2.0.1-1) was also installed from Cydia to locate the directories created by U1Files. The directory name for U1Files in our case was EDF4B87E-CBC0-466C-2377A089DB10. The U1Files directory had five sub directories named Documents, Library, StoreKit, tmp, and U1Files.app. Ubuntu One's default directory was located at the following path:

*/var/mobile/Applications/EDF4B87E-CBC0-466C-2377A089DB10/Documents/Ubuntu One*

We determined that all synced files were stored in that directory. The "Documents" directory contained the "u1.db" database, which we found contains the following three tables:

1- login_info_table (see Figure 21): These appear to be the authentication token components stored by the mobile app for authenticating to Ubuntu One services. The topic of authentication token use in mobile apps is discussed further in Martini, Do and Choo (2015a, 2015b).

| Table Contents | | | | |
|---|---|---|---|---|
| id | consumer_key | consumer_secret | token | token_secret |
| 1 | mHiBm4w | wVFiKYcQSIdFytMSNGfEeQmUSEWUEP | kdEYALSLiecsKWMufBnYJcpNIDsWShZHYvPQvNVFrPdnDRkf | ERYAtMidazPTMHyikfYNbooOJmisAsmlTmPeQhBkNkQRIaL... |

Figure 21. u1.db login_info_table

2- nodes_table (see Figure 22): Filenames, last modified date, size and hash values are the notable values we located in this table.

| Table Contents | | | | | | | | |
|---|---|---|---|---|---|---|---|---|
| parent_path | type | path | content_path | name | is_public | public_url | size | last_modified | hash |
| /~/ | 1 | /~/Ubuntu One | /content/~/Ubu... | Ubuntu One | 0 | | -1 | | |
| /~/Ubuntu One | 2 | /~/Ubuntu One... | /content/~/Ubu... | AQUA-OS2.BMP | 0 | | 153674 | 2014-05-29T03... | sha1:93e06221... |
| /~/Ubuntu One | 2 | /~/Ubuntu One... | /content/~/Ubu... | HANGING.DOC | 0 | | 22016 | 2014-05-29T03... | sha1:7065040f... |
| /~/Ubuntu One | 2 | /~/Ubuntu One... | /content/~/Ubu... | HANGING.txt | 0 | | 2019 | 2014-05-29T03... | sha1:6a1583fe2... |

Figure 22: u1.db nodes_table information

3- version_table (see Figure 23): The currently installed version of the U1Files app can be retrieved from this table. We also located a file named "iTunesMetadata.plist" which contained information including the date that we purchased the U1Files app and the Apple ID that was associated with the purchase.

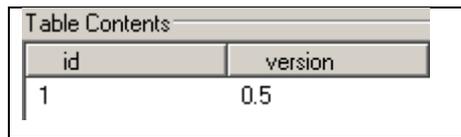
Figure 23: version_table

The results of our network traffic capture analysis on iOS did not differ significantly from our findings for Mac OS X.

## 6. Conclusion

Cloud storage has attracted many individuals and business users by offering cost effective storage and services across a variety of devices. However, the prevalence of cloud storage has provided criminals with opportunities, including the ability to organise their activities in a distributed, scalable and somewhat anonymous way. The nature of the cloud environment imposes several challenges to the traditional process of digital forensic investigation. Data of interest may be segregated on shared storage in different physical locations, where such data is subject to foreign laws and regulations (Hooper, Martini & Choo 2013).

One area of difficulty in cloud forensics is in identifying the use of cloud services by suspects and, then determining the particular cloud service used and acquiring the account credentials for the service (Quick, Martini & Choo 2014). The evidence sources in a digital investigation vary and may include computer hard disks and live memory, network traffic captures and mobile devices such as the Apple iPhone and Android mobile devices. The identification and collection of suspect's data must be carried out in timely fashion before the data can be moved to another unknown location or even permanently deleted. The legislative process for seizing data also differs between jurisdictions. Hence, forensic investigators and practitioners need to preserve suspects data as soon as possible after identification.

In this chapter, we carried out a number of experiments on Ubuntu One to examine the artefacts that may be left behind on client devices after use of this cloud service. Firstly, Windows 8.1 was utilised to access Ubuntu One via its online interface using four major browsers (Internet Explorer, Mozilla Firefox, Apple Safari, and Google Chrome). We then deployed Ubuntu One's client application to Windows 8.1, Apple Mac OS X 10.9 and an unofficial Ubuntu One client on iOS 7.0.4. During our experiments on these platforms, we simulated a user carrying out common operations on cloud storage, such as uploading, downloading and deleting files, with a standard dataset. The evidence sources under investigation varied between on operating systems, but included sources such as the

Windows registry, event logs, network traffic captures, live memory captures and persistent file system changes. We found that it was likely that practitioners would be able to locate a range of distinct remnants in relation to Ubuntu One usage, but access to valuable evidential artefacts (such as authentication and user action logs) varied between platforms. The artefacts described in this chapter may assist forensic practitioners in detecting Ubuntu One use and hopefully assist in acquiring data of potential evidential value.

Cloud storage is likely to remain popular and even grow in usage as the dominant technology used in file hosting and transmission of files among individuals and organisations. As a result, it is recommended that those undertaking future work in this area continue to examine cloud storage including investigation of other cloud storage applications using various devices and operating systems. Future work also includes further analysis of the authentication token system used by the Ubuntu One client, which may allow for the location of authentication credentials on the PC operating systems, in addition to the token credentials located on the mobile client.

### References

Aminnezhad, A., Dehghantanha, A., Abdullah, M., and Damshenas, M. (2013) Cloud Forensics Issues and Oppurtunities. *International Journal of Information Processing and Management,* 4(4),76-85.

Biggs, S., and Vidalis, S. (2009). Cloud Computing: The Impact on Digital Forensic Investigations. In *International Conference for Internet Technology and Secured Transactions,*.

Birk D, W. C. (2011). Technical issues of forensic investigations in cloud computing environments. In *6th international workshop on systematic approaches to digital forensic engineering – IEEE/SADFE 2011,* (pp. 1–10).

Chung, H., Park, J., Lee, S., & Kang, C. (2012). Digital forensic investigation of cloud storage services. *Digital Investigation*, 9(2), 81–95.

Do, Q., Martini, B., Looi, J., Wang, Y., & Choo, K.-K. R (2014). Windows Event Forensic Process. *IFIP International Federation for Information Processing*, 433, 87-100.

Daryabar, F., Dehghantanha, A., & Udzir, N. I. (2013). A Review on Impacts of Cloud Computing on Digital Forensics. International Journal of Cyber-Security and Digital Forensics (IJCSDF), 2(2), 77-94.

Federici, C. (2014). Cloud data imager: A unified answer to remote acquisition of cloud storage areas. Digital Investigation, 11(1), 30-42.

Grispos, G., Storer, T., & Glisson, W. (2012). Calm before the storm: The challenges of cloud computing in digital forensics. *International Journal of Digital Crime and Forensics*, 4(2), 28-48.

Hale, J. S. (2013). Amazon Cloud Drive forensic analysis. *Digital Investigation*, 10(3), 259–265.

Hooper, C., Martini, B., & Choo, K. K. R. (2013). Cloud computing and its implications for cybercrime. *Computer Law & Security Report*, 29(2), 152-163.

Martini, B., & Choo, K. K. R. (2013). Cloud storage forensics: ownCloud as a case study. Digital Investigation, 10(4), 287-299.

Martini, B., & Choo, K.-K. R. (2012). An integrated conceptual digital forensic framework for cloud computing. *Digital Investigation*, 9(2), 71–80.

Martini, B., Do, Q., & Choo, K.-K. R (2015a). Conceptual evidence collection and analysis methodology for Android devices. In Ko R and Choo K-K R, editors, Cloud Security Ecosystem, Syngress, an Imprint of Elsevier

Martini, B., Do, Q., & Choo, K.-K. R (2015b). Mobile cloud forensics: An analysis of seven popular Android apps. In Ko R and Choo K-K R, editors, Cloud Security Ecosystem, Syngress, an Imprint of Elsevier

McKemmish, R (1999). What is Forensic Computing?. Trends & Issues in Crime and Criminal Justice, 118, 1 - 6.

Mee, V., Tryfonas, T., & Sutherland, I. (2006). The Windows Registry as a forensic artefact: Illustrating evidence collection for Internet usage. *Digital Investigation*, 3(3), 166-173.

Mell, P., & Grance, T. (2011). The NIST Definition of Cloud Computing-Recommendations of the National Institute of Standards and Technology. NIST. *NIST Special Publication*.

Damshenas, M., Dehghantanha, A., Mahmoud, R., & bin Shamsuddin, S. (2012, June). Forensics investigation challenges in cloud computing environments. In Cyber Security, Cyber Warfare and Digital Forensic (CyberSec), 2012 International Conference on (pp. 190-194). IEEE.

Quick, D., & Choo, K.-K. R. (2013a). Digital droplets: Microsoft SkyDrive forensic data remnants. *Future Generation Computer Systems*, 29(6), 1378–1394.

Quick, D., & Choo, K.-K. R. (2013b). Dropbox analysis: Data remnants on user machines. *Digital Investigation*, 10(1), 3–18.

Quick, D., & Choo, K.-K. R. (2013c). Forensic collection of cloud storage data: Does the act of collection result in changes to the data or its metadata? *Digital Investigation*, 10(3), 266–277.

Quick, D., & Choo, K.-K. R. (2014). Google Drive: Forensic analysis of data remnants. *Journal of Network and Computer Applications*, 40, 179–193.

Quick, D., Martini, B., & Choo, K.-K. R. (2014). *Cloud Storage Forensics*. Syngress Publishing / Elsevier.

Taylor, M., Haggerty, J., Gresty, D., & Lamb, D. (2011). Forensic investigation of cloud computing systems. *Network Security*, 2011(3), 4–10.